\documentclass[letterpaper]{article}
\pdfoutput=1 %
\usepackage[utf8]{inputenc}
\usepackage{amsthm,amsfonts}
\usepackage{algorithm}
\usepackage{algpseudocode}
\usepackage{aas_macros}
\usepackage{hyperref} %
\usepackage{csquotes} %
\usepackage[backend=biber,style=numeric]{biblatex}
\usepackage{subfiles}

\newcommand{\J}{Jonathan\xspace}%

\PassOptionsToPackage{textwidth=3cm}{todonotes}%
\usepackage{contraction-tree}

\renewcommand{\t}[1]{\tensortype{#1}} %
\newcommand{\A}{\t{A}} %
\newcommand{\B}{\t{B}} %

\renewcommand{\lblname}{\cutsetname}%
\newcommand{\alt}{\widehat}%
\NewSubscriptFunc{\hd}{\alt{\lblname}}

\newcommand{\contreeHeight}{2in} %

\newcommand{\nodes}{\Vpar}

\renewcommand{\defeq}{=}
\newcommand{\ieq}{=}

\toggletrue{preferSubscriptsToParens} %
\togglefalse{showOptionalSubscripts} %
\toggletrue{preferFreeToUnrooted}

\title{Carving-width and contraction trees for tensor networks}

\author{J. Jakes{-}Schauer, D. Anekstein, P. Wocjan}
\renewcommand{\a}[1]{%
  	\ifnumcomp{#1}{=}{1}{a}{%
    	\ifnumcomp{#1}{=}{2}{a'}{%
        	\ifnumcomp{#1}{=}{3}{a''}{%
            	\ifnumcomp{#1}{=}{4}{a'''}{%
                  \csname Bad arg to macro. \endcsname
                }
            }
        }
    }
}%

\newcommand{\bagname}{b}
\newcommand{\bagI}[1]{\bagname(#1)}
\newcommand{\bag}[1]{%
	\ifnumcomp{#1}{=}{0}{%
    	\bagname 
    }{
    	\ifnumcomp{#1}{=}{1}{%
          \bagI
        }{%
          \csname ERROR: Bad arg to \bag macro. \endcsname
        }%
    }%
}%

\renewcommand{\T}[1]{#1}%
\renewcommand{\cTf}{\mathcal{T}}%

\renewcommand{\cTd}{\mathcal{T}_{\mathrm{d}}} %

\newcommand{\mapname}{\phi}
\renewcommand{\n}{|\V{}|}

\begin{document}

\commenting{G.2.3 [Mathematics of Computing]: Graph Theory---Applications; G.2.2 [Mathematics of Computing]: Graph Theory---Graph algorithms, Trees; J.2 [PHYSICAL SCIENCES AND ENGINEERING]: Physics}

\maketitle

\abstract{%
We study the problem of finding contraction orderings on tensor networks for physical simulations using a syncretic abstract data type, the $\emph{contraction-tree}$, and explain its connection to temporal and spatial measures of tensor contraction computational complexity (nodes express time; arcs express space).  We have implemented the Ratcatcher of Seymour and Thomas for determining the carving-width of planar networks, in order to offer experimental evidence that this measure of spatial complexity makes a generally effective heuristic for limiting their total contraction time.%
}%

\section{Introduction}
\label{sec:Background and intro}
Tensors, for current purposes, are functionally multidimensional arrays of complex numbers.  A \emphdef{tensor network} is a factorization of a large tensor into a family of simpler tensors which tracks their interrelationships--usually graphically, as by \maybehref{https://en.wikipedia.org/wiki/Penrose_graphical_notation}{Penrose notation}--so as to make the uncompressed tensor product recoverable in principle.  
Good introductions are freely available (e.g.~\cite{Practical-intro, Nutshell, Hand-waving}) into the role of the tensor network in models of quantum chemistry, quantum many-body physics, and, most recently, classical approximations of quantum computation.  In these simulations, the contents of the component tensors stand for superpositions of quantum particles, and their bonds for the degrees of entanglement between them.  A maximally entangled system, of course, is not simplified by factorization; but the importance of locality in the systems being modeled enables it in practice.  With this approach one can manipulate features of tensors whose unfactored form would require computational resources %
 far in excess of what is physically possible, not to mention practical.  

The pinch is felt when a calculation calls for the scalar equivalent of a closed network, as this requires contracting all bonds, i.e., indices; the complexity of which procedure is usually exponential in the number of component tensors.  If the operation known as `contraction' is the generalization of the trace of a two-dimensional tensor, then contracting two or more tensors is a generalization of \maybehref{https://en.wikipedia.org/wiki/Matrix_chain_multiplication}{matrix chain multiplication}; like with matrices, the order of the pairwise contractions is irrelevant to the value obtained but highly consequential for the efficiency.  Full contraction of even minimally entangled networks has been proved to be \maybehref{https://en.wikipedia.org/wiki/Sharp-P}{$\sharp$P-hard}~\cite{Sharp-P} and generally unfeasible, providing strong motivation to identify the exceptional cases.  Our present preoccupation, the problem of finding a best contraction order for an existing tensor network---as opposed to incorporating different factorizations into the search---is historically known %
as \emphdef{single-term optimization}, has a clean graph-theoretic analogy, and is unfortunately NP-hard~\cite{LamLoops}, in contrast to the one-dimensional matrix-multiplication problem.%

Optimization, for matrix multiplication or tensor contraction, generally means minimizing the total number of paired multiplication-addition operations in a multiply-nested loop, approximating, in a big-Theta sense, the time required by a single-core processor in the standard random-access machine (RAM) computation model.\commenting{Is that the right name for that ``machine"?}  Any further compile-time or runtime accelerations can improve on this strategy only up to a hardware-dependent multiplicative constant.  This sequential-time-complexity metric is the one used with \Netcon~\cite{Netcon}, which is still, to our knowledge, the most aggressive graph-based single-term optimization algorithm, but which, by dint of its exactitude, is usable only for small to moderately sized networks of less than one hundred tensors.  
In fact, rather than focus primarily on contraction order, it is much more common to use lossy compression techniques~\cite{Survey-of-contraction-methods}.  
An exception is the well-known MPS method~\cite{MPS}, which is unidimensional enough that finding and executing a contraction order can be managed efficiently.  However, one has only to look at the bidimensional version of MPS, PEPS~\cite{PEPS}, to see both tasks reach maximum difficulty.

Our interest is in networks with many tensors, but low average tensor dimension, for which an exact contraction result is required but an approximately minimal contraction \emph{cost} is acceptable. 
This kind of result could be used, for instance, to validate the approximate contraction methods currently in vogue.  
Correspondingly, an approximation technique that minimizes the cost, not of the entire contraction, but of the most expensive part of it, leads us to carving-width, which, introduced in~\cite{Ratcatcher}, is one of a family of ``widths'' researched at length by Robertson and Seymour for quantifying edge complexity of a graph.  
The original of these, and the most studied, is %
\maybehref{https://en.wikipedia.org/wiki/Treewidth}{\emphdef{treewidth}};
its applicability to tensor networks was noted in an influential paper by \textcite{Markov-Shi}, used practically by \textcite{Breaking49}, and benchmarked comparatively by \textcite{ConSequence}.
On the other hand, the relevance of carving-width to tensor networks was, to our knowledge, first remarked upon in an arXival e-print by \textcite{bubble-width}, who, rediscovering it in a
restricted form, knew it as `bubble width'; it is used in its primary sense by \textcite{OliveiraOliveira2017}.
We have come to believe that its obscurity is undeserved, for several reasons:
\begin{enumerate}
\item  The (comparatively) well-known treewidth was designed for unweighted graphs, so it only works on tensors of uniform bond dimension.  %
This may hold for quantum circuits, but not other species of tensor networks, which is probably why it is has seen more use in circuit simulation than in the other quantum disciplines.  A generalized weighted treewidth~\cite{WeightedTreewidth} is extant, but has not received the same level of attention.  By contrast, carving-width works with all nonnegative real edge weights.%
\nolinebreak\footnote{Seymour and Thomas used natural numbers, but the extension to reals is uncomplicated, up to the imprecision of floating-point calculation.}  
\item Second, because treewidth requires a vertex-weighted graph, not edge-weighted, one must take the \maybehref{https://en.wikipedia.org/wiki/Line_graph}{line graph} first.  
As shown in~\cref{sec:carving-width}, finding a \emphdef{carving-decomposition} automatically yields a tree-decomposition of the line graph.  
\item Most importantly, the carving-decomposition constructed for a minimal carving-width is a binary tree and provides the basis for a datatype, explained in the next section, which best describes an arbitrary contraction order.  In other words, the step between assessing the cost of a near-optimal contraction order, and finding the order itself, is short.
\end{enumerate}

An obstacle to using them, even as a means to a larger end, is that carving-width~\cite{Ratcatcher} and treewidth (including for unweighted graphs~\cite{Treewidth-is-NP-complete}) are both NP-complete problems %
 and must themselves be approximated.  For the important class of planar graphs 
(drawn with no crossing edges), however, there exists the original Seymour-Thomas algorithm, rather whimsically called the `\ratcatcher,' which can find the carving-width in pseudopolynomial time; whereas %
2019 has passed with the existence of a comparable exploit for treewidth remaining an open problem, providing one more motivation for working with carving-width.  %
In \cref{sec:Application} we demonstrate how finding the carving-decomposition of a tensor network (using the \ratcatcher algorithm) can provide %
reasonable contraction orders for said network, provided that it is planar, and we demonstrate an application to a PEPS-like grid.  %
\Cref{sec:conclusion} discusses potential extensions.
\section{Graph-theoretical formulation%
}
\label{sec:graph-theory-defs}
\label{sec:Preliminaries}
A tensor network is, conveniently, isomorphic to an undirected weighted graph $G\ieq\p{\V0,\E0,w}$ without self-loops or parallel edges.  Vertices are identified with tensors, edges with tensor indices, and edge weights with index dimensions; the degrees of the vertices give the tensors' orders.  Edges are identified with subsets of $\V0$ of cardinality $2$.
  Parallel edges, while supportable, are disallowed, because any bundle of such edges may be replaced with a single edge, weighted with the product of their weights.  Loops can similarly be eliminated by summing over the corresponding indices~\cite{Netcon}.  Importantly, these simplification operations never harm efficiency by adding to the resource requirements of contraction.  
  Also, there is no semantic distinction between an edge with weight $1$ and a nonexistent edge, meaning that they may be interchanged at will.
\label{no-free-indices}
Finally, a tensor network often has indices initially left \emph{free}, to serve as model inputs.  This does not a valid graph make, so we assume that free indices get bound to probability vectors and contracted preliminarily.%
\nolinebreak\footnote{The alternative, tying all the dangling edges to a new vertex and leaving them uncontracted, increases contraction time exponentially in the number of inputs.}

An example graph $\G$ is given in \cref{fig:2x3 tensor}: a `tensor train' as would be part of a MPS process.
\smallfig{%
\centering
\begin{tikzpicture}
\node[TensorStyle](A) {$\T A$};
\node[TensorStyle, right = of A](B) {$\T B$};
\node[TensorStyle, right = of B](C) {$\T C$};

\node[TensorStyle, below = of A](D) {$\T D$};
\node[TensorStyle, below = of B](E) {$\T E$};
\node[TensorStyle, below = of C](F) {$\T F$};

\draw (A) -- node[BondLabelStyle] {$a$} (B) -- node[BondLabelStyle]{$b$} (C);
\draw (A) -- node[BondLabelStyle] {$c$} (D);
\draw (B) -- node[BondLabelStyle] {$d$} (E);
\draw (C) -- node[BondLabelStyle] {$e$} (F);
\draw (D) -- node[BondLabelStyle] {$f$} (E) -- node[BondLabelStyle] {$g$} (F);

\end{tikzpicture}
\caption{A $2\times 3$-tensor network}
\label{fig:2x3 tensor}
}%
We use capitals $A,B,\ldots,F$ to label vertices and miniscules %
$a,b,\ldots,g$ for edges.  %
For instance, the index joining tensors $A$ and $B$, $a$, is equivalent to the edge $\cb{A,B}$, and has dimension $\wI{a} = \wpar{\cb{A,B}}$, with $\weightname : \E0 \to \Np$ denoting the weight function.

\subsection{Contraction}
\Cref{fig:2x3 minor} shows a possible contraction sequence for the graph of \cref{fig:2x3 tensor}.
\def\@minorSize{0.34\textwidth}%
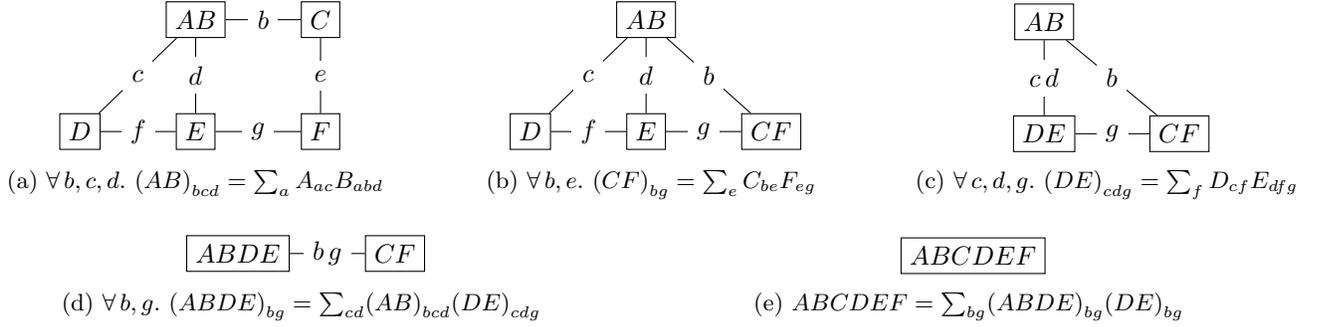
\begin{figure}[ht]
	\subcaptionbox{$\Forall b,c,d.\ \T{\p{AB}}_{bcd} = \sum_a\T{A}_{ac} \T{B}_{abd}$%
    	\label{fig:first-minor}}
    [\@minorSize]
    {\begin{tikzpicture}
\node[TensorStyle](AB) {$\T{AB}$};
\node[TensorStyle, right = of AB](C) {$\T C$};

\node[TensorStyle, below = of AB](E) {$\T E$};
\node[TensorStyle, left = of E](D) {$\T{D}$};
\node[TensorStyle, below = of C](F) {$\T F$};

\draw (D) -- node[BondLabelStyle] {$c$} (AB) -- node[BondLabelStyle]{$b$} (C);
\draw (AB) -- node[BondLabelStyle] {$d$} (E);
\draw (C) -- node[BondLabelStyle] {$e$} (F);
\draw (D) -- node[BondLabelStyle] {$f$} (E) -- node[BondLabelStyle] {$g$} (F);

\end{tikzpicture}}
	\subcaptionbox{$\Forall b,e.\ \T{\p{CF}}_{bg} = \sum_e\T{C}_{be} \T{F}_{eg}$}
    [\@minorSize]
    {\begin{tikzpicture}
\node[TensorStyle](AB) {$\T{AB}$};

\node[TensorStyle, below = of AB](E) {$\T E$};
\node[TensorStyle, left = of E](D) {$\T{D}$};
\node[TensorStyle, right = of E](CF) {$\T{CF}$};

\draw (D) -- node[BondLabelStyle] {$c$} (AB) -- node[BondLabelStyle]{$b$} (CF);
\draw (AB) -- node[BondLabelStyle] {$d$} (E);
\draw (D) -- node[BondLabelStyle] {$f$} (E) -- node[BondLabelStyle] {$g$} (CF);

\end{tikzpicture}}
	\subcaptionbox{$\Forall c,d,g.\ \T{\p{DE}}_{cdg} = \sum_f \T{D}_{cf} \T{E}_{dfg}$%
    	\label{fig:third-minor}}
    [\@minorSize]
    {\begin{tikzpicture}
\node[TensorStyle](AB) {$\T{AB}$};

\node[TensorStyle, below = of AB](DE) {$\T{DE}$};
\node[TensorStyle, below right = of AB](CF) {$\T{CF}$};

\draw (DE) -- node[BondLabelStyle] {$c\,d$} (AB) -- node[BondLabelStyle]{$b$} (CF);
\draw (DE) -- node[BondLabelStyle] {$g$} (CF);

\end{tikzpicture}}
\par\bigskip
	\subcaptionbox{$\Forall b,g.\ \T{\p{ABDE}}_{bg} = \sum_{cd} \T{\p{AB}}_{bcd} \T{\p{DE}}_{cdg}$}
    [0.5\textwidth]
    {\begin{tikzpicture}
\node[TensorStyle](ABDE) {$\T{ABDE}$};

\node[TensorStyle, right = of ABDE](CF) {$\T{CF}$};

\draw (ABDE) -- node[BondLabelStyle] {$b\,g$} (CF);

\end{tikzpicture}}
	\subcaptionbox{$\T{ABCDEF} = \sum_{bg} \T{\p{ABDE}}_{bg} \T{\p{DE}}_{bg}$}
    [0.5\textwidth]%
    {\begin{tikzpicture}
\node[TensorStyle](ABCDEF) {$\T{ABCDEF}$};

\end{tikzpicture}}
    \caption{A succession of graph minors of~\cref{fig:2x3 tensor}, showing a full contraction and the accompanying summations.  Edges are being contracted in the sequence $\sqb{a,\,e,\,f,\,cd,\,bg}$.}
	\label{fig:2x3 minor}
\end{figure}%

We enumerate the tensors that arise during the contraction of a network using subsets of $\V0$.  For any $X\subseteq V$, the tensor corresponding to $X$ is the one formed by merging all the (fundamental) tensors of $X$; that is, by contracting all edges in the induced subgraph $\G[X]$.  In step~\ref{fig:third-minor}, for example, the contraction is of tensors $X = \cb{A,B}$ and $X' = \cb{D,E}$, outputting $X\cup X' = ABDE$.  Clearly, due to the distributive law the output of 
a sequence of contractions does not depend on the order of combination.

  However, the cost, in terms of computational complexity, does.  
To illustrate this we need the concept of a `cut', which is a separation of $\G$ into disconnected subgraphs by the removal of a subset of $\E0$.

\begin{definition}[$2$-cut]%
\label{def:2-cut}
A $2$-cut, \emph{edge-cut}, or just \emph{cut} is a bipartition of $\E0$.
For $X,X'\subseteq \V0$ with $X\cap X'=\emptyset$, define the \emphdef{cut-set} $\cutsetI{X,X'}$ as the set of edges needed to disconnect $G[X]$ from $G[X']$:
\begin{equation}
	\label{eq:def. of 2-cut}
	\cutsetI{X,X'} \defeq \sbuild[big]{ \{A,B\} \in \E }{ A\in X, B\in X' }\,.%
\nolinebreak\footnote{The $\cutsetname$ notation is adapted from~\cite{Ratcatcher}.}
\end{equation}
In other words, $\cutsetI{X,X'}$ is the set of edges having one endpoint in $X$ and one in $X'$. %
\end{definition}

This generalizes easily to the idea of a multiway $m$-cut $\cutsetI{X_1,\ldots,X_m}$; but the only other case we will specifically need is $m=3$:
\begin{definition}[$3$-cut]
\label{def:3-cut}
For $X,X',X''\subseteq V$ with $X\cap X'=\emptyset$, $X\cap X''=\emptyset$, and $X'\cap X''=\emptyset$, define
\begin{equation} \label{eq:def. of 3-cut}
	\cutsetI{X,X',X''} \defeq \cutsetI{X,X'} \dunion \cutsetI{X,X''} \dunion \cutsetI{X',X''}\,.%
\nolinebreak\footnote{We use $\dunion$ to emphasize that this is a union of disjoint sets.}
\end{equation}
\end{definition}%

For subsets of edges $F\subseteq \E0$, it useful to define the weight of $F$ to be the product of the weights of the edges in $F$, that is,
\begin{equation*}
\wpar{F}=\prod_{e\in F} \w{e}\,.
\end{equation*}

This makes possible:
\begin{definition}[Cut-weight]
For $X_1,\ldots,X_m$ with 
$\bigdunion_i X_i = \V0$, the \emphdef{cut-weight}, $\wpar{X_1,\ldots,X_m}$, is given by
\begin{equation*}
\wpar{X_1,\ldots,X_m} = \wpar{\cutsetI{X_1,\ldots,X_m}}\,,
\end{equation*}
with obvious applicability to $2$- and $3$-cuts.
\end{definition}
To motivate use of these definitions, we first need to refine the meaning of `contraction sequence.'
We start the contraction sequence $\cS$ with the $\n$ fundamental tensors as input, identifying each of which with a singleton subset of $\V0$.
For every contractive step in the sequence, we always select two tensors $X$ and $Y$ and contract them to a new tensor $Z=XY$. In total, we create $\n-1$ new tensors by pairwise contractions, with the last one being the scalar value of the fully contracted network.  %
Each tensor created along the way corresponds to the union of the preceding subsets of $\V0$---mapping, in turn, to tensors earlier in the sequence.  We now have a list of $2\n-1$ subsets of $V$, with the second half describing the contractions.  E.g., for \cref{fig:2x3 minor}, 
\begin{gather*}
    \cS = \Bigl[
    \cb{A}^1,\cb{B}^2,\cb{C}^3,\cb{D}^4,\cb{E}^5,\cb{F}^6,
    \Bigr. \\
    \Bigl. \cb{A,B}^7, \cb{C,F}^8, \cb{D,E}^9, \cb{A,B,D,E}^{10}, \cb{A,B,C,D,E,F}^{11}
    \Bigr]\,,
\end{gather*}
\renewcommand{\P}{P^{(i)}}%
\newcommand{\GP}{\G_{\P}}%
with positional IDs added as superscripts.  Each contractive step $i\in \ii{\n}{2\n-1}$ %
can also be associated with a \maybehref{https://en.wikipedia.org/wiki/Graph_minor}{minor} of $\G$, that is, a partially contracted graph, as follows.
Let $\P$ be the partition of $\V0$ formed by taking the first $i$ elements of the sequence, removing any member which is a subset of another.  $\P$ is homomorphic to the minor $\GP \ieq \p{\P,\E{\P},\wsub{\P}}$, where, for all $X,Y\in \P$,
\begin{equation*}
	\weight[\P]{X,Y} \defeq \weight[\G]{X,Y} = \wpar{\cutsetII{X,Y}{\G}}\,;
\end{equation*}
while $\E{\P}$ is simply the set of vertex pairs for which $\wsub{\P}$ is not 1.%
Generalizing these precedential relationships, we see that each $\P$ is a layer in a \maybehref{https://en.wikipedia.org/wiki/Semilattice}{join-semilattice}.  More specifically, a \emphdef{contraction order} can be described by a binary tree %
 $T$ with an injective map
$\mapname :\ii{1}{\p{2\n-1}} \to \V{T}$.
This $\mapname$ must satisfy the so-called \maybehref{https://en.wikipedia.org/wiki/Binary_heap}{max-heap} property: the ID of each internal node has to be greater that the IDs of both its children.  Intuitively, an ID marks the point in $\cS$ at which the corresponding tensor is produced, and it makes sense that its two constituent tensors must have been created first.  Thus the vertices of $\G$ map to $T$'s leaves.

\bigskip

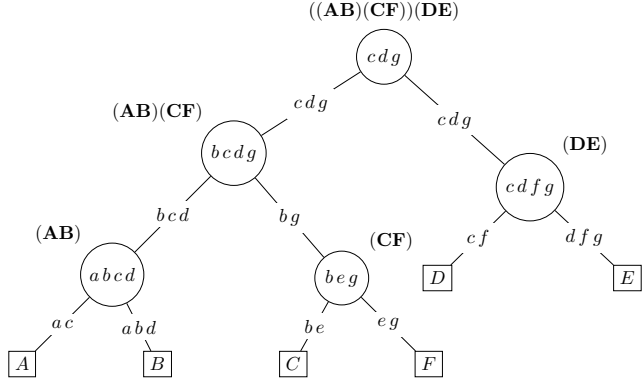
\begin{figure}%
\centering
\resizebox*{!}{\contreeHeight}{\newcommand{\Clab}[1]{\mathbf{#1}}%
\renewcommand{\T}{\Clab}%

\begin{tikzpicture}%
\node[TensorStyle] (A) {$A$};
\node[TensorStyle, right = 2 of A] (B) {$B$};
\node[TensorStyle, right = 2 of B] (C) {$C$};
\node[TensorStyle, right = 2 of C] (F) {$F$};

\node[BagStyle, BondLabelStyle, above right = of A] (AB) {$a\,b\,c\,d$};
\node[BagStyle, BondLabelStyle, above left = of F] (CF) {$b\,e\,g$};

\node[TensorStyle, right = of CF] (D) {$D$};
\node[TensorStyle, right = 3 of D] (E) {$E$};

\node[BagStyle, BondLabelStyle, above right = 2 of AB] (ABCF) {$b\,c\,d\,g$};
\node[BagStyle, BondLabelStyle, above right = of D] (DE) {$c\,d\,f\,g$};

\node[BagStyle, BondLabelStyle, above right = 1 and 2 of ABCF] (ABCDEF) {$c\,d\,g$};

\node[ContractionLabelStyle, above left = 1pt of AB] (AB-label) {$\T{\p{AB}}$};
\node[ContractionLabelStyle, above right = 1pt of CF] (CF-label) {$\T{\p{CF}}$};
\node[ContractionLabelStyle, above left = 1pt of ABCF] (ABCF-label) {$\T{\p{AB}\p{CF}}$};
\node[ContractionLabelStyle, above right = 1 pt of DE] (DE-label) {$\T{\p{DE}}$};
\node[ContractionLabelStyle, above = 1pt of ABCDEF] (ABCDEF-label) {$\T{\p{\p{AB}\p{CF}}\p{DE}}$};

\draw (A) -- node[BondLabelStyle] {$a\,c$} (AB) -- node[BondLabelStyle] {$a\,b\,d$} (B);
\draw (C) -- node[BondLabelStyle] {$b\,e$} (CF) -- node[BondLabelStyle] {$e\,g$} (F);
\draw (AB) -- node[BondLabelStyle] {$b\,c\,d$} (ABCF) -- node[BondLabelStyle] {$b\,g$} (CF);
\draw (ABCF) -- node[BondLabelStyle] {$c\,d\,g$} (ABCDEF) -- node[BondLabelStyle] {$c\,d\,g$} (DE);
\draw (D) -- node[BondLabelStyle] {$c\,f$} (DE) -- node[BondLabelStyle] {$d\,f\,g$} (E);
\end{tikzpicture}}
\caption{Example rooted \ctree for \cref{fig:2x3 tensor}, with additional labels in boldface.}
\label{fig:contraction-binary-tree}
\end{figure}

Formally:
\begin{definition}[Rooted \ctree]%
\label{def:rooted ctree}
A rooted \ctree $\cTr$ of the weighted graph $G \ieq \p{\V0,\E0,\weightname}$ is a rooted, labeled, unordered, full binary tree with each external node identified with a unique $v\in \V0$.
This tree type was used earlier in%
~\cite{Oliveira}, though the labeling offered there differs somewhat from ours.  We label both nodes and arcs,%
\nolinebreak\footnote{To reduce ambiguity, we call components \emph{vertices} and \emph{edges} when referring to the network $\G$ and \emph{nodes} and \emph{arcs} only with \ctrees---a small abuse of terminology for undirected trees.}
 as follows:
\begin{itemize}
\item
The removal of an arc $a$ partitions the leaves into the two disjoint sets $X_a$ and $\overline{X}_a$; the label of arc $a$ is 
\[
\lbl_a \defeq \cutsetI{X_a,\overline{X}_a}\,.
\]
\item
The removal of an internal node $n$ partitions the leaves into the three mutually disjoint sets $X_n$, $Y_n$, and $\overline{X_n Y_n}$; the label of node $n$ is %
\[
\lbl_n \defeq \cutsetI{X_n, Y_n, \overline{X_n Y_n}}\,.
\]
For the special case of the root node, one of these sets is empty, making it a %
 de facto 2-cut.  The other internal nodes are all 3-cuts.
\item 
The label of an external node $v$ is
\[
\lbl_v \defeq \sbuild{ e  \in \E }{ v\in e }\,,
\]
that is, all edges incident at $v$. Note that $\lbl_v$ is equal to $\lbl_a$, where $a$ is the unique arc connecting the leaf node to the rest of the \ctree.  Due to the redundancy, labels of this sort will be omitted from figures, such as \cref{fig:contraction-binary-tree}.  %
\end{itemize}
\begin{indention}
\begin{remark}
It is important to observe that the nodes $n$ of the rooted \ctree $\cTr$ could be naturally identified with subsets of $\V0$. However, we do not make this identification a required part of the definition because---as we shall shortly see---it will be very useful to consider `free' \ctrees with which such identification is no longer possible.
\end{remark}
\end{indention}
\end{definition}%

It should be clear that many contraction sequences give rise to the same $\cTr$, per the variability of the $\mapname$ mapping.
The benefit of using 2- and 3-cuts %
(\cref{def:2-cut,def:3-cut}), is that they provide a complexity measure which abstracts away unnecessary information. 

\begin{definition}[Tensor space complexity]
The number of bytes needed to store tensor $X$, up to a multiplicative constant, is the cut-weight
\begin{equation*}
	\wpar{X,\overline{X}}\,,
\end{equation*}
and for a \ctree arc $a$ with $\lbl_a = \cutsetI{X_a,\overline{X}_a}$, 
\begin{equation*}
	\w{a} \defeq \wpar{\lbl_a} = \wpar{X_a,\overline{X}_a}\,.
\end{equation*}
\end{definition}%
\begin{definition}[Contraction time complexity]
Symmetrically, the number of primitive operations in contracting tensors $X$ and $Y$ to get $XY=X\dunion Y$ is, up to a multiplicative constant, the cut-weight
\begin{equation*}
	\wpar{X,Y,\overline{XY}}\,.%
\end{equation*}
For a \ctree node $n$,
\begin{equation*}
	\w{n} \defeq \wpar{\lbl_n} = \wpar{X,Y,\overline{XY}}.
\end{equation*}
\end{definition}%

\begin{definition}[Space bottleneck]
Define the \emphdef{space bottleneck},
 $\Bspace{\cTr}$, of a \ctree as the maximum over all arcs:
\[
\Bspace{\cTr} = \max_{a} \cb[big]{ \w{a} }\,.
\] 
\end{definition}%

\begin{definition}[Time bottleneck]
Symmetrically, define the \emphdef{time bottleneck}, $\Btime{\cTr}$, as the maximum over all nodes:
\[
\Btime{\cTr} = \max_{n} \cb[big]{ \w{n} } \,.
\] 
For unweighted graphs, this reduces to the \emphdef{contraction complexity} of \textcite{Markov-Shi}.%
\end{definition}%

\begin{definition}[Total time]
Finally, define the \emphdef{total time complexity}, $\Ctime{\cTr}$, the sum over all internal nodes:
\[
\Ctime{\cTr} = \sum_{n} \w{n}\,.
\]
\end{definition}%

It is trivially true that $\Bspacename \le \Btimename \le \Ctimename$, a relationship we will examine further in the next section.  For now, note that given the root $r$ with incident arcs $\a1$ and $\a2$, then 
$\lbl_r = \lbl_{\a1} = \lbl_{\a2}$.  This means that the root node never determines $\Bspacename$ or $\Btimename$, suggesting that we consider a homeomorphic rootless tree.

\begin{definition}[{[Free] \ctree}]
An \emphdef{unrooted} or \emphdef{free \ctree} (in future, just \emphdef{\ctree}) $\cTf$ of the weighted graph $\G \ieq \p{\V0,\E0,\weightname}$ is a free, unordered binary tree with leaves mapped bijectively to $\V0$ and nodes and arcs labeled as in \cref{def:rooted ctree}, save, of course, the root $r$.  
Any $\cTr$ can be converted to a $\cTf$ by removing the root and splicing its arcs together so as to make its children neighbors.  $\Bspacename$ and $\Btimename$ are the same for $\cTf$ and $\cTr$, while $\Ctime{\cTr} = \Ctime{\cTf} + \w{r}$.  A contraction sequence can only be built from a rooted tree; however, because we can always ``re-root'' $\cTf$ optimally (and efficiently; see \cref{sec:postprocessing}), the contribution made by $r$ to the time complexity is asymptotically negligible.  %
   Hence we will treat all \ctrees as free by default.%
\begin{figure}%
\centering
\resizebox{!}{\contreeHeight}{%
	\begin{tikzpicture}
\node[BagStyle, BondLabelStyle] (p-root) {$b\,c\,d\,g$};

\node[BagStyle, BondLabelStyle, above left = of p-root](AB) { $a\,b\,c\,d$ };
\node[TensorStyle, above  = of AB](A) {$\T A$};
\node[TensorStyle, below left = of AB](B) {$\T B$};

\node[BagStyle, BondLabelStyle, above right = of p-root](CF) {$b\,e\,g$};
\node[TensorStyle, above  = of CF](C) {$\T C$};
\node[TensorStyle, below right = of CF](F) {$\T F$};

\node[BagStyle, BondLabelStyle, below = of p-root](DE) {$c\,d\,f\,g$};
\node[TensorStyle, below left = of DE](D) {$\T D$};
\node[TensorStyle, below right = of DE](E) {$\T E$};

\draw (A) -- node[BondLabelStyle] {$a\,c$} (AB) -- node[BondLabelStyle]{$a\,b\,d$} (B);
\draw (C) -- node[BondLabelStyle] {$b\,e$} (CF) -- node[BondLabelStyle]{$e\,g$} (F);
\draw (AB) -- node[BondLabelStyle] {$b\,c\,d$} (p-root) -- node[BondLabelStyle] {$b\,g$} (CF);
\draw (D) -- node[BondLabelStyle] {$c\,f$} (DE) -- node[BondLabelStyle] {$d\,f\,g$} (E);
\draw (p-root) -- node[BondLabelStyle] {$c\,d\,g$} (DE);

\end{tikzpicture}
}%
\caption{\Ctree on \cref{fig:2x3 tensor} (\iftoggle{preferFreeToUnrooted}{`free'}{rootless} version of \cref{fig:contraction-binary-tree}).}
\label{fig:contraction-tree}
\end{figure}
\end{definition}%

\section{Space %
\textit{vs.} time optimization}
\label{sec:treewidth and carving-width}
$\Bs0$, $\Bt0$, and $\Ct0$ for \ctrees measure local optima; their definitions can be made global with respect to a given $\G$ simply by minimizing over all possible \ctrees.  There are $\p{2\n-3}!!$ such $\cTf$.%
\nolinebreak\footnote{I.e., the number of free, unordered, terminally-labeled binary trees~\cite{counting-rooted-contraction-trees}.  The rooted kind has $\p{2\n-5}!!$.}  %
  The \Netcon algorithm of \textcite{Netcon} uses branch-and-bound techniques to construct %
$\argmin\,\Ctime{\cTr}$ %
directly; however, it turns out that $\Bspace{\G}$ and $\Btime{\G}$ also are the targets of existing algorithmic methods.  The choice of labels for the \ctree was made to match this intuition: informally, arc labels denote space complexity, and node labels, time; and either can be obtained from its counterpart.  
\cref{sec:treewidth} and \cref{sec:carving-width} provide the formal details as to how the \ctree encodes, respectively, treewidth and carving-width, while \cref{sec:Relationships} offers some simple bounds.

\subsection{Weighted treewidth and \texorpdfstring{$\Btimename$}{Bt}}
\label{sec:treewidth}
A \emphdef{tree-decomposition} of a graph, also known as a \emphdef{junction tree}, \commenting{as described in~\cite{Treewidth}} is formed by grouping the vertices of that graph into sets, or `bags,' according to a trifecta of rules---which are reviewed below---so that these bags form the nodes of a tree.  The object is to keep the bags as small as possible, because the \emphdef{width} of the tree-decomposition is determined by the cardinality of the largest bag; the \emphdef{treewidth}, or $\tw{\G}$, is the smallest width of any tree-decomposition feasible; %
and a variety of algorithms that are NP-hard for arbitrary graphs---tensor networks, for example---are exponential merely in the treewidth, becoming tractable when that is provably bounded.  For weighted graphs, the extension to a \emphdef{weighted treewidth}, or $\wtw{\G}$, by \textcite{WeightedTreewidth} is more appropriate.  

However, because either form of treewidth is defined for vertex-weighted graphs---whereas ours have weighted edges---we must operate instead on the \emphdef{line graph}%
, a construction in which vertices map to members of $\E0$ and cliques to neighbors of members of $\V0$.

\begin{definition}[Weighted tree-decomposition of $L$]
The weighted tree-decomposition $\cTd$ of the line graph $L$ of $\G=\p{\V0,\E0,\weightname}$ 
 is a tree whose nodes $n$ are labeled by subsets $\bag1{n}$ of $\E0$, together %
satisfying the following properties:
\begin{enumerate} \label{tree-decomp props}
\item \label{enu:property 1 of a tree decomposition}
For each index $i\in \E0$, there exists at least one node $n$ with $i\in \bag1n$.

\item \label{enu:property 2 of a tree decomposition}
For each pair of indices $i'$ and $i''$ incident on the same vertex $v\in \V0$, there exists at least one node $n$ with $\cb{i',i''} \subseteq \bag1n$.

\item \label{enu:property 3 of a tree decomposition}
For each index $i\in \E0$, the subgraph induced by the subset of nodes $\sbuild{n}{i\in \bag1n}$ is connected.  In the machine-learning literature this is known as the \emph{running intersection property}.%
\end{enumerate}
The \emphdef{weighted width} of the tree-decomposition $\cTd$ is 
\[
\ww{\cTd} \defeq \max_n \Big\{ \wpar{b_n} \Big\}\,,
\]
where the maximum is taken over all nodes of the tree $\cTd$.

The weighted treewidth of the line graph, $\wtw{L}$, is the minimal weighted width
\[
	\wtw{L} \defeq \min_{\cTd} \Big\{ \ww{\cTd} \Big\}\,,
\]
taken over all admissible tree-decompositions of $L$.
\end{definition}
Instead of counting the bags' contents, one multiplies the associated weights.  The original application was to \maybehref{https://en.wikipedia.org/wiki/Bayesian_network}{Bayesian networks}, where the weight $\weightII{i}{\G}$ measured the domain of some random variable; here we use it as the range of a tensor index.  
\begin{theorem}[\Ctree $\mono$ tree-decomposition]
An arbitrary \ctree $\cTf$, modulo %
\footnote{Stripping the leaf nodes is optional.  They will never contribute to the treewidth.}%
 its 2-cuts $\cutsetname_a$, forms a tree-decomposition of the line graph $L$ having width $\Btime{\cTf}$.
\end{theorem}
\begin{proof}
Let $e \ieq \cb{v,v'}$ be any index in $\E0$.  By construction of the labels of the leaf nodes, we have $e\in\labI{v}$ and also $e\in\labI{v'}$.  
Properties~\ref{enu:property 1 of a tree decomposition} and~\ref{enu:property 2 of a tree decomposition} follow immediately.  

This leaves the running intersection property.

\begin{nestedproof}
\begin{lemma}[\RIP] %
	\label{lem:path property}
Let $\cTf$ be an arbitrary free \ctree. For any $e\in \E0$, the subgraph induced  by the subset of nodes 
\[
\sbuild{n}{e\in \labI{n}}
\]
is a path.
\end{lemma}
\begin{proof}

Begin with a version of $\cTf$ which is ``bare,'' having leaf nodes still mapped to $\V0$ but being otherwise unlabeled.   Then construct an alternative labeling $\alt\lblname$ using the desired property and show that the resulting tree $\alt\cTf$ is equivalent to $\cTf$.  
The alternative labeling $\hd{n}$ of the nodes and $\hd{a}$ of the arcs is defined thus:

\begin{itemize}
\item For every leaf node $v$, set $\hd{v}=\edges{v}$.
\item For every internal node $n$ and every arc $a$, add $e\in \E0$ to their label sets $\hd{n}$ and $\hd{a}$, respectively, if and only if they are on the unique path connecting the two leaf nodes %
 $v$ and $v'$, with $e\in \edges{v}\cap \edges{v'}$.
\end{itemize}

It suffices to show that $\cTf$ and $\alt\cTf$ coincide on all labels.  For the external nodes this is immediate.  For any internal node $n$, recall that the definition of a 3-cut %
(\cref{eq:def. of 3-cut}) partitions the leaf set into nonempty subsets $X \dunion Y \dunion Z = V$, with 
\begin{equation*}
	\labI{n} = \cutsetI{X,Y,Z} = \cutsetI{X,Y} \dunion \cutsetI{X,Z} \dunion \cutsetI{Y,Z}\,.
\end{equation*}
Meanwhile, in $\alt\cTf$, the \RIP---explicitly supported---means that $i$ appears in $\hd{n}$ if and only if it has its endpoints in exactly two of $\cb{X,Y,Z}$:
\begin{equation*}
\newcommand{\xor}{\lor}
  \Forall e\, \Forall n.\ e\in\hd{n} \iff e\in\cutsetI{X,Y} \xor e\in\cutset{X,Z} \xor e\in\cutset{Y,Z}\,,
\end{equation*}
formally equivalent to the 3-cut definition.   A similar argument applies to arc labels in $\alt\cTf$ and the definition of 2-cut %
(\cref{eq:def. of 2-cut}).  $\cTf$ and $\alt\cTf$ are the same up to monomorphism.

\begin{remark}
The information contained within the node labels %
${\lblname_n}$ %
of a \ctree is the same as that compassed by the arc labels 
${\lblname_a}$, in the sense that from either set the other may be deduced.  Using the running intersection property, it is not hard to prove that, for some arc $a\ieq\cb{n,n'}$, 
\begin{equation}\label{eq:intersection-property}
	\labI{a} = \labI{n} \cap \labI{n'}\,,
\end{equation}
or that for any internal node $n$, 
\begin{equation}
    \label{eq:union-property}
	\labI{n} = \labI{a} \cup \labI{a'}
\end{equation}
where $a$ and $a'$ are two of its three arcs.%
\end{remark}%
\end{proof}%
\end{nestedproof}%
This finishes property~\ref{enu:property 3 of a tree decomposition}. 
\end{proof}%

\begin{theorem}[Tree-decomposition $\epi$ \ctree] 
Let $\cTd$ be an arbitrary tree-decomposition of the line graph $L$ of $\G$.  Then, $\cTd$ can always be efficiently transformed to a new tree-decomposition $\cTd'$ such that $\ww{\cTd'}\le\ww{\cTd}$ and $\cTd'$ forms a free \ctree.
\end{theorem}
\begin{proof}
Use the following %
three-step %
algorithm to meet the specific structural requirements of a \ctree (degree-3 internal nodes, and so on):

\newcommand{\hTd}{\alt{\cT}_{\mathrm{d}}}
\begin{enumerate}
\item \label{step:add-leaves}
As before, take the bare version of $\cTd$ from which all but the leaf labels have been struck. Grow the tree by attaching some $\n$ new leaf nodes identified with $v\in V$. The attachment point for each $v$ may be an arbitrary $n$ such that $\edgesI{v}\subseteq \bag1{n}$; for example, the lexicographic minimum.  
Calling this new tree $\hTd$, empty all its internal nodes' bags entirely and refill them  from $\V0$ based on the alternative labeling used in \cref{lem:path property}, that is, by putting $e = \cb{v,v'}$ into the path between $v$ and $v'$.  Label the arcs correspondingly.

$\hTd$ with bags $\hd{n}$ at this point defines a valid tree-decomposition of $\L{G}$: properties \ref{enu:property 1 of a tree decomposition} and \ref{enu:property 2 of a tree decomposition} are trivially satisfied by the leaf nodes $v$, where we have $\hd{v}=\edges{v}$.
Property~\ref{enu:property 3 of a tree decomposition} is satisfied because the labeling $\hd{n}$ meets the stronger condition that the subgraph $\hTd\sqb{\sbuild[big]{v}{ e\in \hd{v}}}$ induced, for any edge $e\in \E0$, is a path.  
Moreover, because $\hd{n}\subseteq \bag1{n}$, thanks to the \RIP, we know that the weighted width, $\ww{\hTd}$, cannot have grown.

All nodes left with empty bags should be contracted into their neighbors---as should nodes with degree $2$.  (This sub-step is unnecessary if we can stipulate from the start that no bag in $\cTd$ is a subset of another, a condition known in e.g.~\cite{Reduced-tree-decomps} as \emphdef{reduced} form.)  Any remaining degree-$1$ node must be one of the newly added leaves.
    
\item Label arcs using \cref{eq:intersection-property}.  $\hTd$ is now a free \ctree, except that its internal nodes may not be---indeed, likely are not---ternary.

\item \label{step:split-bags}
\renewcommand{\m}[1]{%
	\ifstrequal{#1}{1}{%
    	m'%
    }{%
    	\ifstrequal{#1}{2}{%
        	m''%
        }{%
        	\csname ERROR in 'm' macro\endcsname%
        }
    }%
}
Each node with more than three arcs now gets split, with the process repeated until all that remain have degree 3.  Let $n$ be an offending node, and let $\a2, \a3$ be any two of its arcs.  We \emphdef{split} $n$ into two nodes $\m1, \m2$ and connect them with a new arc $\a4$ so that 

\begin{align*}
	\edges{\m1} &= \cb{\a2, \a3, \a4} \\
    \edges{\m2} &= \cb{\a4} \dunion \p{\edges{n} \setminus \cb{\a1,\a2}}%
\end{align*}
with labels
\begin{align*}
	\hd{\m1} &= \hd{\a2} \cup \hd{\a3} \\
	\hd{\m2} &= \bigcup \, \sbuild[Big]{\hd{a}}{a \in \edges{\m2}} \\
	\hd{\a4}   &= \hd{\m1} \cap \hd{\m2}\,.
\end{align*}
Note that this is the same labeling scheme used in \cref{lem:path property}.
Clearly, the widths of the bags of $\m1$ and $\m2$ cannot be larger than that of $n$; $\card{\edges{\m1}} =3$ and $\card{\edges{\m2}} = \card{\edges{n}} - 1$.

\begin{remark}
If desired, it is always possible to find a split of $n$ so that at least one of $\m1$ or $\m2$ has strictly smaller width.   
In nontrivial cases, that is, unless $\Exists v\in \V0$ such that $\hd{n} = \edges{v}$, the splitting process can be used as an opportunity to shrink both child nodes.
\end{remark}
\end{enumerate}

At this point, we have a free binary tree with $\cardinality{\nodes{\hTd}}=2\n-2$ whose external nodes correspond one-to-one to the sets $\sbuild{\edges{v}}{ v\in\V0}$ . The resulting node- and arc-labeled tree forms a \ctree, up to epimorphism.  The entire process of trimming and relabeling can be done with, naïvely, time complexity in $\O{\card{\V0} \cdot \card{\E0}}$.
\end{proof}

\subsection{Carving-width and \texorpdfstring{$\Bspacename$}{Bs}}
\label{sec:carving-width}
Introduced by \textcite{Ratcatcher} as an ancilla to a third tree-based metric, \emphdef{branchwidth}, 
carving-width is the least well known of the three.   
Weighted-graph branchwidth and carving-width are notable for their own merit in addressing the `call-routing problem' for telecoms. %
We will omit branchwidth from the discussion and proceed directly to carving-width%
.  One prompt advantage over treewidth is that
the edges-to-vertices interchange in the last section is needed no longer.%
\begin{definition}
\label{def:carving-decomp}
\newcommand{\Tc}{\mathcal{C}}
A \emphdef{carving-decomposition}, also known as a \emphdef{routing tree}, %
 $\Tc$
of $\G$ is a free, full binary tree supporting the vertices of $\G$ for leaves.  
For each $a \in \E{\Tc}$, removing $a$ would partition $\V{\G}$ into two sets, $S_1,S_2$, by the remaining connected components.  Each $a$ is labeled with the cut-edges $\cutset{S_1,S_2}$.  The \emphdef{load} of $a$ is the sum of weights 
$\sum \sbuild{\w{e}}{e\in \cutsetI{S_1,S_2}}$; %
the \emph{width} or \emphdef{congestion} of $\Tc$ is the heaviest load in the tree; and the \emphdef{carving-width}, $\carw{\G}$, is defined as the lowest achievable width, symmetrically to treewidth.%
\nolinebreak\footnote{The frankly more appealing $cw\p{\G}$ is often seen, but that abbreviation can also mean \emphdef{cut-width}, another tree complexity measure.}
\end{definition}
\begin{theorem}[Free \ctree $\iso$ carving-decomposition]\label{thm:carv contr}
An arbitrary \ctree $\cTf$, modulo its 3-cuts $\cutsetname_n$, is isomorphic to a carving-decomposition.
$\Bspace{\cTf}$ is equal, up to logarithmic concavity, to the carving-width:
\begin{equation*}
	\carw{\G} = \log{\Bs{\G}}\,.
\end{equation*}
\end{theorem}
\begin{proof} Straightforwardly, the 2-cuts of the \ctree duplicate, by design, the branch structure of a carving-decomposition.  The isomorphism between weights is merely a matter of transferring between the groups $\p{\N,+}$ and $\p{\R^{*},\times}$.
\end{proof}

\subsection{\texorpdfstring{$\Bspacename$}{Bs}, \texorpdfstring{$\Btimename$}{Bt}, and \texorpdfstring{$\Ctimename$}{Ct} united}
\label{sec:Relationships}

\begin{theorem}
\label{thm:Bs Bt Ct}
  The asymptotic relationship between $\Bs0$, $\Bt0$, and $\Ct0$ is fairly tight:
\begin{gather}
	\Bs0 \le \Bt0 \le \p{\Bs0}^{1.5} \label{eq:Bs-Bt} \\
    \Btimename + 4\p{\n-3} \le \Ctimename \le \p{\n-2}\Btimename \, \label{eq:Bt-Ct}
\end{gather}
whether taken with respect to a given \ctree or its graph.
\end{theorem}%

\begin{proof}
The left inequality of \cref{eq:Bs-Bt}, $\Bs0 \le \Bt0$, is trivial.  
For the right, use the following lemma:

\renewcommand{\a}[1]{%
  	\ifnumcomp{#1}{=}{1}{a}{%
    	\ifnumcomp{#1}{=}{2}{a'}{%
        	\ifnumcomp{#1}{=}{3}{a''}{%
            	\ifnumcomp{#1}{=}{4}{a'''}{%
                  \csname Bad arg to macro. \endcsname
                }
            }
        }
    }
}%

\begin{nestedproof}
  \begin{lemma} For any internal node $n$ in a free \ctree with adjacent arcs $\a1,\a2,\a3$,
  \begin{align}
  	\w{n} = \sqrt{\w{\a1}\, \w{\a2}\, \w{\a3}}\,.
  \end{align}
  \end{lemma}
  \begin{proof}
  By the \maybehref{https://en.wikipedia.org/wiki/Inclusion\%E2\%80\%93exclusion_principle}{inclusion-exclusion principle} of measure theory,
  \[
  	\w{n} = \w{\a1}\, \w{\a2}\, \w{\a3}
     \ \div\ 
    	\p{%
        	\prod_{e \in \lab{\a1} \cap \lab{\a2}} \w{e}
    		\prod_{e \in \lab{\a1} \cap \lab{\a3}} \w{e}
    		\prod_{e \in \lab{\a2} \cap \lab{\a3}} \w{e}
		}%
     \ \times\ 
     	\prod_{e \in \lab{\a1} \cap \lab{\a2} \cap \lab{\a3}} \w{e}\,.
  \]
Because our networks are not hypergraphs, the three-way intersection $\lab{\a1} \cap \lab{\a2} \cap \lab{\a3}$ is always empty.  Using set algebra, we can rewrite the large product that is the middle expression as

\[
 \prod\, \sbuild[big]{\w{e}}{ e\in \p{\lab{\a1} \cap \lab{\a2}}
 \dunion \p{\lab{\a1} \cap \lab{\a3}}
 \dunion \p{\lab{\a2} \cap \lab{\a3}} }
\]
\[
 \prod\, \sbuild[Big]{\w{e}}{ e\in 
 	\sqb[big]{ \lab{\a1} \cap \p{\lab{\a2} \cup \lab{\a3}} }
 \cup 
 	\sqb[big]{ \lab{\a2} \cap \p{\lab{\a1} \cup \lab{\a3}} }
 \cup
 	\sqb[big]{ \lab{\a3} \cap \p{\lab{\a1} \cap \lab{\a2}} }
 }%
 \,.
\]

Recalling that ${\lab{\a2} \cup \lab{\a3}} = {\lab{\a1} \cup \lab{\a3}} = {\lab{\a1} \cap \lab{\a2}} = \lab{n}$, from \cref{eq:union-property}, this reduces to
\begin{gather*}
 \prod\, \sbuild[big]{ \w{e} }{ e\in 
 	\p{ \lab{\a1} \cap \lab{n} } 
 \cup 
 	\p{ \lab{\a2} \cap \lab{n} }
 \cup
 	\p{ \lab{\a3} \cap \lab{n} }
 }%
 \\
\prod\, \sbuild[big]{\w{e}}{ e \in \p{\lab{\a1} \cup \lab{\a2} \cup \lab{\a3}} \cap \lab{n} } \\
	\prod_{e\in \lab{n}} \w{e}
\end{gather*}

which is simply $\w{n}$.  Thus
$
	\w{n} = {\w{\a1}\, \w{\a2}\, \w{\a3}} \div {\w{n}}
$
as was intended.
  \end{proof}
\end{nestedproof}%
Now assume $\w{n}=\Bt1{\cT}$.  Knowing that $\a1,\a2,$ and $\a3$ cannot exceed $\Bs1{\cT}$, then in the worst case we have $\Bt1{\cT} = \sqrt{\p{\Bs1{\cT}}^3}$, which finishes \cref{eq:Bs-Bt}.
The same is true for $\Bs1{G}$ and $\Bt1{G}$ because $\Bs1{\cT}$ cannot be worse than $\p{\Bs1{\cT'}}^{1.5}$ for the $\cT'$ with the best possible carving-width.

The inequalities of \cref{eq:Bt-Ct} are merely the minimum and maximum possible when summing over all $\n-2$ contractions.  In the case of the lower bound, the minimal $\ctimename$-value for an individual contraction is $4$.  This is because, in order for $\G$ to be connected, at most one out of the three 2-cuts which form $\cutsetI{n}$ may be empty, with cut-weights $\ge 2$ for the other two.
\end{proof}
\begin{remark}
It is not guaranteed that a $\Bspacename$-optimal \ctree should be $\Btimename$-optimal, or conversely.  \Cref{fig:Bs-Bt-mismatch} shows a simple $K_4$ graph for which the \ctree indicated by $\cb{AB,CD}$ is $\Bspacename$-optimal, but that of $\cb{AC,BD}$ is $\Btimename$-optimal.
\smallfig{%
\centering
\resizebox*{!}{1in}{%
	\begin{tikzpicture}
\def\a{6}
\def\b{4}
\def\c{4}
\def\d{3}
\def\e{2}
\def\f{2}

\node[TensorStyle](A) {$\T{A}$};
\node[TensorStyle, right = 2 of A](B) {$\T B$};

\node[TensorStyle, below = 2 of A](C) {$\T C$};
\node[TensorStyle, below = 2 of B](D) {$\T D$};

\node[below right = 0.25 of A] (AD) {$\e$};
\node[below left = 0.3 of B] (BC) {$\f$};

\draw (A) -- node[BondLabelStyle] {$\a$} (B);
\draw (A) -- node[BondLabelStyle] {$\b$} (C);
\draw (A) -- (AD) -- (D);
\draw (B) -- (BC) -- (C);

\draw (B) -- node[BondLabelStyle] {$\c$} (D);

\draw (C) -- node[BondLabelStyle] {$\d$} (D);

\end{tikzpicture}%
}
\caption{$\Bspacename$ and $\Btimename$ uncorrelated.  Edge labels are bond dimensions.}
\label{fig:Bs-Bt-mismatch}
}%
\end{remark}

\section{Efficient computation of contraction orders for planar tensor networks} 
\label{sec:Application}
\subsection{`Ratcon'}
The \ratcatcher algorithm, which gets its name from a game theory analogy the authors of~\cite{Ratcatcher} use in their verification proof,  addresses what is referred to as the call-routing problem: given a collection of calls made between a set of locations, design a network to route these calls such that the maximum network congestion is minimized. Provided the graph representing calls between locations is planar, Seymour and Thomas demonstrated that the decision problem of whether the maximum congestion, the carving-width, is less than some integer $k$ can be solved in polynomial time. %
The \ratcatcher is used as a subroutine in incrementally constructing a carving-decomposition for an input graph. Knowing that carving-width and $\Bspacename$ are analogues under the isomorphism outlined in \cref{thm:carv contr}, it is possible to find the minimal $\Bspacename$ and a corresponding free \ctree for a planar tensor network in polynomial time. In this section we review our implementation of the \ratcatcher algorithm and its use in constructing a rooted \ctree with the corresponding carving-width, from which a space-optimal contraction order is derived. We will refer to the entire process as \emph{\ratcon}.

\subsection{Generation of sample planar tensor networks for experimentation}
\label{sec:experimental-design}
We describe now how we chose the tensor networks used to evaluate the performance of the \ratcon algorithm. We wanted our tensor networks to resemble those that occur naturally when describing and simulating interesting quantum many-body systems.

\emphdef{Matrix product states} (MPS; e.g.~\cite{MPS}) play a popular role in simulating one-dimensional quantum systems.  Because the network graph of such a system is a path, a MPS tensor can be contracted very efficiently by treating all contractions as matrix multiplications. \emphdef{Projected entangled pair states} (PEPS; e.g.~\cite{PEPS}) are a two-dimensional generalization of MPS %
whose subsystems correspond to the vertices of triangular, square, or hexagonal grids.  Unfortunately, PEPS tensor networks cannot be contracted efficiently, and it is typically necessary to perform approximate instead of exact contractions when faced with simulating a system of more than several dozens of tensors.  %
Singular value decomposition is the tool employed in these `truncation' methods to reduce the bond dimensions of tensors that grow too large~\cite{Survey-of-contraction-methods}.
The goal of the present paper is understanding the computational cost when the contraction phase, at least, is as exact as possible.  We would like to have an efficient algorithm for computing a close-to-optimal contraction ordering for PEPS networks, in order to bring the cost of exactitude down to the realm of the possibility for instances containing hundreds of tensors.  %

\begin{figure}[ht]
\begin{center}
\includegraphics[width=2in]{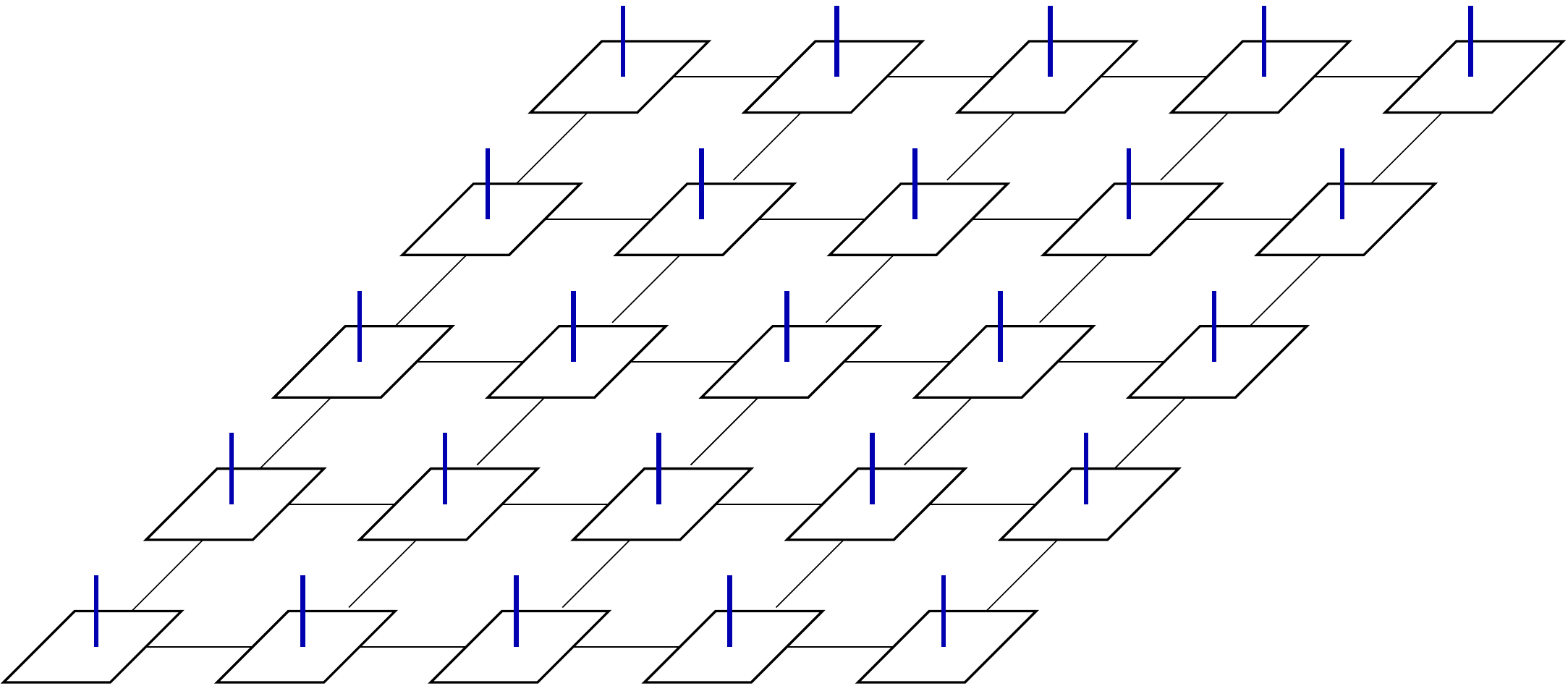}
\end{center}
\caption{Ket tensor network for a PEPS state.}
\label{fig:half_sandwich}
\end{figure}

\begin{figure}[ht]
\begin{center}
\includegraphics[width=2in]{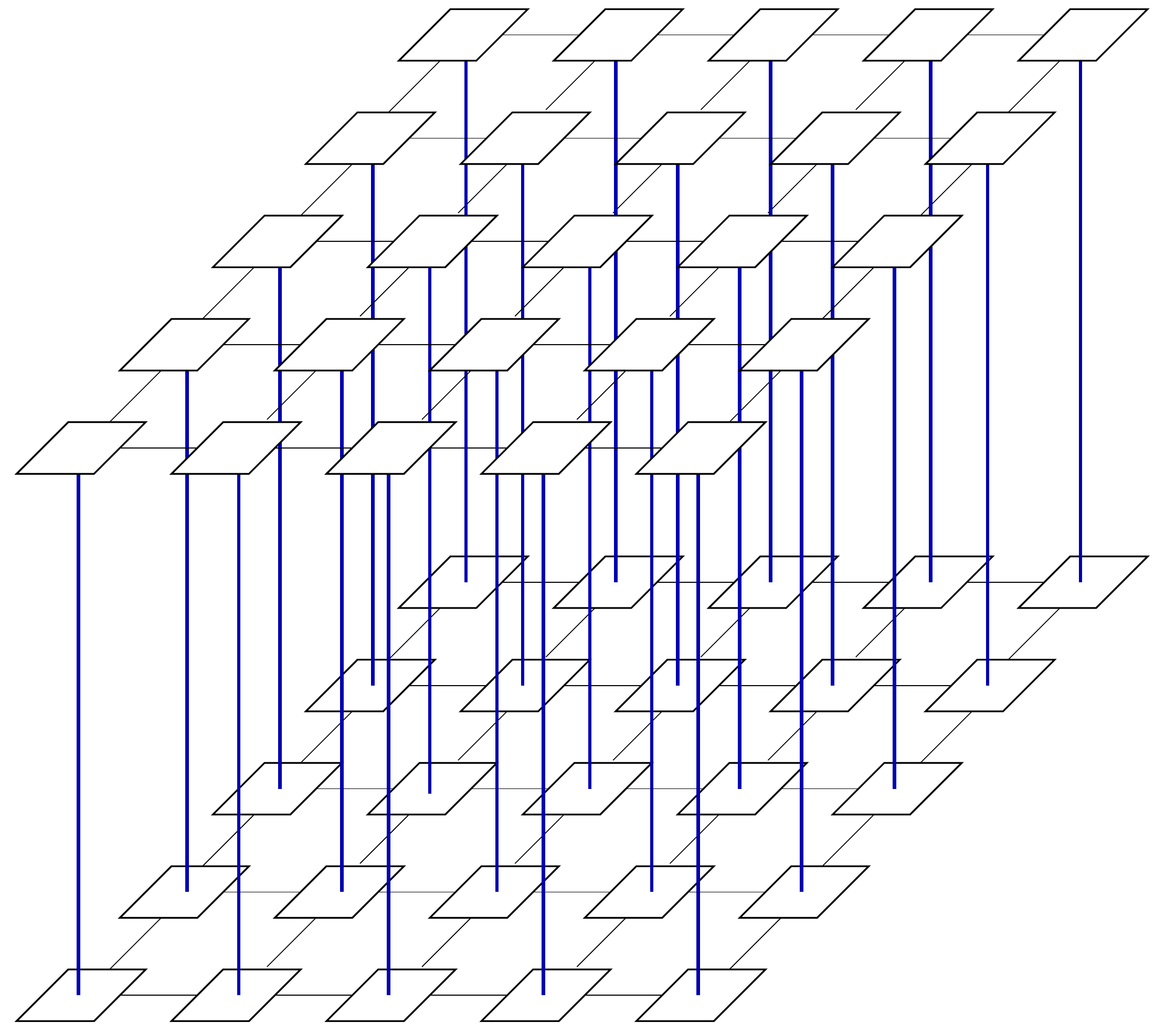}
\end{center}
\caption{Bra-ket tensor network for a PEPS state.}
\label{fig:sandwich}
\end{figure}

\Cref{fig:half_sandwich} shows a PEPS two-dimensional quantum state consisting of qubits arranged on a $5 \times 5$ grid. The perpendicular edges are free indices of weight $2$ (because the state space of a qubit is a two-dimensional Hilbert space). The weights of the horizontal edges correspond to so-called bond dimensions and are in general nonuniform.
\Cref{fig:sandwich} shows a tensor network of a shape that often needs contracting to compute some interesting physical quantity, such as the expected value of an observable. 

It is clear that this ``bra-ket'' graph is not planar and, thus, that it is not possible to directly apply the \ratcon algorithm.  Na\"{i}ve ways of planarizing these graphs, such as ``squashing'' the upper and lower grids by prioritizing contraction of the vertical edges, tend to produce highly objectionable scores, because the dimensions of the virtual bonds get squared.  
To test the performance of the \ratcon algorithm, then, we have settled for considering half-graphs, that is, square grids that look like the graph in \cref{fig:half_sandwich}, but without the free edges.  Tests on the full PEPS forms are deferred until such time as an approximation extension into three dimensions can be found.

\subsubsection{Lognormal prior}
To find a domain which produces challenging, but not impossible, samples, we begin with square grid graphs---because of the relation to PEPS, and because of their high genera, with which they are naturally complex---and apply independent bond dimensions as edge weights, subject to acceptance-rejection testing.  We wish to make as few assumptions about prior probabilities as possible.  %
To restrict the space to  ``realistic'' sizes, we reject any sample graph which would require more than 5\,TiB of working memory, assuming 16-byte complex-valued tensor elements: chosen to reflect the amount required by \textcite{Breaking49} in their exact simulation.%
\nolinebreak\footnote{The precise value they report is $2^{42}$ bytes.}  %
  Thus $\Bspace{\G} \le 2^{36}$.
Furthermore, to rule out insufficiently entangled systems, we keep only graphs which would remain biconnected after removal of unit edges.

A plain uniform distribution being useless under these restrictions, we use instead a normal prior---or rather lognormal, because every sample thereof can be rounded to a positive integer, and which should better reflect the exponential growth that occurs over the course of, for example, imaginary time evolution~\cite{ITE}.%
\nolinebreak\footnote{Our colleagues provide in \cite{Simulation-on-AWS} a more detailed explanation of ITE applied to PEPS, and how it motivated the search for better contraction heuristics.}  %
The mean $\mu$ for the lognormal distribution, or rather $\exp(\mu)$, varies with $L$, set to the largest value for which a uniform grid would be contractible within the aforestated memory bounds.  The standard deviation $\sigma$ is allowed to vary from edge to edge, as a uniform, continuous, i.i.d. random variable.  %
  Maxima for $\sigma$ are determined, for each $(L,\,\mu)$ pair in turn, by estimating, through a simple examination of a graphical plot of many candidates' $\Bspacename$ scores, where the probability of finding a graph meeting the feasibility criteria would drop, effectively, to zero.  Only an upper bound is needed.

The object for this sample population, as filtered through the rejection criteria, is to represent graphs at the extreme of what can be done practically and without loss, where the contraction order will make the greatest difference.

\subsection{Pre-processing of tensor networks%
}\label{sec:Ratcon}
Once a tensor network for a given $L$ has had its bond dimensions generated, the logarithmic mapping described in \cref{thm:carv contr} is applied to the network. Unlike the original integral formulation of carving-width for the \ratcatcher, we now have real-valued, or rather floating-point, weights.  This reduces the accuracy of the carving-width representation according to the arithmetic constraints of the architecture used. 

Following the logarithmic mapping, the carving-width of the input graph is calculated using a modified binary search: the lower bound of this search  begins at the largest fundamental bond dimension, and the upper bound is determined by incrementally and exponentially scaling a value $k$ until the \ratcatcher reports that it has exceeded the carving-width.  %
A conventional binary search is then conducted on the interval $[\nicefrac{k}{2}, k)$ until the carving-width is found.  The time complexity of the \ratcatcher component, which solves the decision problem at each iteration, is $\O{\n^2}$.%
\nolinebreak\footnote{As $\G$ is planar, $\E\G \in \Th{\V\G}$.}  The \ratcatcher is called $\Th{\log_2 \BsI{\G}}$ times in this binary search and so runs in $\O{\n^2\cdot \log\BsI{\G}}$ time. The algorithm is pseudopolynomial due to the dependency of carving-width on the total weight of the graph, and not the number of edges.  %
Importantly, this proportionality to $\carw\G = \log_2 \BsI{\G}$ should only pose a problem when $\Ctimename$ is so large as to prohibit contraction---in which case the whole exercise would be pointless in practice.

With the carving-width now available, a carving-decomposition%
\nolinebreak\footnote{Actually a `carving', a set-theory isomorph of the carving-decomposition.} %
 can be constructed using what is called by %
 \textcite{HicksRat,HicksCycle} the \emphdef{edge-contraction} algorithm, which incrementally locates %
 eligible edges of the partially-contracted graph: %
 an edge is \emphdef{eligible} if its weight is no more than %
 the carving-width of the current minor; if its contraction results in a minor that is also biconnected; and if this new minor has carving-width no more than that of the original $\G$. %

 To improve contraction total time, the edge-contraction procedure must be done more than once, because a typical graph supports many $\Bspacename$-optimal \ctrees with widely varying $\Ctimename$ values.  A simple compensation is to perform many random edge contractions and keep the best one.
This makes the time complexity for this phase of processing $\O{N\cdot \n^4\cdot \log\BsI{\G}}$, where $N$ is chosen for the edge-contraction sample size.
 
\subsection{Post-processing of carving-decompositions}
\label{sec:postprocessing}
The carving-decomposition returned by the edge-contraction phase is next used to generate a space-efficient contraction order. This step can be broken into two: rooting the carving-decomposition (free \ctree) to yield a rooted \ctree, which provides a contraction template with a locally-optimal time bottleneck, and then deriving a contraction order from that template.  When rooting a free \ctree, the goal is to minimize the resulting $\Ctimename$. Finding a location to root the tree is slightly dependent on whether the tensor network will be contracted sequentially or in parallel, i.e., whether multiple contractions can be done simultaneously.
        
Here we treat only the most straightforward case of a \emph{sequential time-optimal strategy}, which means inserting a root node $r$ %
into $\cTf$ so that $\CtimeI{\cTr}$ is minimized.  Since $\Ctime{\cTr} - \Ctime{\cTf} = \wI{r}$, we have only to minimize the complexity $\wI{r}$ itself, which we can do very quickly by splitting the edge %
        $$ \argmin_{e\in\E\cTf}\w{e} $$        
and naming that new node as the root.
\newcommand{\Troot}{\texttt{node}\xspace}
\newcommand{\Tleft}[1]{\ensuremath{\leftchild{\text{#1}}}\xspace}
\newcommand{\Tright}[1]{\ensuremath{\rightchild{\text{#1}}}\xspace}
\newcommand{\leftseq}{\texttt{lseq}\xspace}
\newcommand{\rightseq}{\texttt{rseq}\xspace}
\newcommand{\lspace}{\ensuremath{\cspacename_{\ell}}}
\newcommand{\rspace}{\ensuremath{\cspacename_{r}}}
\newcommand{\leftCs}{\ensuremath{\Cspacename_{\ell}}\xspace}
\newcommand{\rightCs}{\ensuremath{\Cspacename_{r}}\xspace}
\newcommand{\lcost}{\texttt{lfirst}\xspace}
\newcommand{\rcost}{\texttt{rfirst}\xspace}
\newcommand{\contraction}{\texttt{contraction}}
\newcommand{\seq}{\texttt{seq}\xspace}
\begin{algorithm}[ht]
\caption{Memory-optimization heuristic}\label{alg:mem-seq}
\begin{algorithmic}
\Procedure{Sequence}{\Troot}
\If{\Troot is a leaf}
	\State \Return \big($\emptylist$, $\Cspace{\Troot}$, $\Cspace{\Troot}$\big)
\Else
	\State \big(\leftseq, \leftCs, \lspace\big) $\gets$ \Call{Sequence}{\Tleft{\Troot}}
    \State \big(\rightseq, \rightCs, \rspace\big) $\gets$ \Call{Sequence}{\Tright{\Troot}}
    \State \lcost $\gets$ \lspace + \rightCs \Comment{Cost of contracting the left subtree first}
    \State \rcost $\gets$ \rspace + \leftCs \Comment{Cost of contracting the right subtree first}
    \State \contraction $\gets$ $\{\Tleft{\Troot}$, $\Tright{\Troot}\}$ \Comment{Identify the subtree contracted next}
    \If{\lcost $\le$ \rcost}
    	\State \seq $\gets$ \leftseq \concat \rightseq \concat [\contraction] \Comment{Contract left subtree first}
        \State \Return \Big(\seq, $\max \big\{\cspaceI{\Troot}$, \lcost\big\}, $\cspaceI{\Troot}$\Big)
    \Else
    	\State \seq $\gets$ \rightseq \concat \leftseq \concat [\contraction] \Comment{Contract right subtree first}
        \State \Return \Big(\seq, $\max \big\{ \cspaceI{\Troot}$, \rcost\big\}, $\cspaceI{\Troot}$\Big)
    \EndIf
\EndIf
\EndProcedure
\end{algorithmic}
\end{algorithm}

Secondarily, we conserve memory in the process of generating a contraction sequence.  We can now offer a measure of space complexity $\CspacenameG$, pursuant to the assumption that the resources for storing a tensor may be reclaimed as soon as that tensor has been contracted with another, but not before the contraction has finished, and that the multiplicands and their product must be stored separately.  This is not strictly optimal, because it always fully contracts one of the subtrees before attending to the other, but is adequate for a proof of concept.%
\nolinebreak\footnote{An always-optimal algorithm is described by \textcite{Memory-optimal-evaluation}.}

Let $\Tr$ be a rooted \ctree.  For convenience, define $\cspaceI{n}$ for every node.  	If $n$ is an endpoint of arc $a = \cb{n,n'}$, where $n$ is farther from the root---deeper---than $n'$, then let
$$ \cspaceI{n} \defeq \wI{a} $$
with $\cspaceI{r} = 1$.  If $n$ is an internal node, we designate its child nodes as $\leftchild{n}$ and $\rightchild{n}$.  Then 
$$\CspaceI{\Tr} \defeq \CspaceI{r}$$
using the recursive definition
\begin{customeq}
    	\CspaceI{n} &\defeq
        	\begin{dcases*}
            	\wI{n} & \ensuremath{\where n \in \V{\G}} \\
                \max\left\{
                	\begin{array}{l}
                    	\cspaceI{n}, \\
                        \min\left\{
                        	\begin{array}{c}
                            	\CspaceI{\leftchild{n}} + \cspaceI{\rightchild{n}}, \\
                            	\cspaceI{\leftchild{n}} + \CspaceI{\rightchild{n}}
                            \end{array}
                        \right\}
                    \end{array}
                \right\} & otherwise.%
            \end{dcases*}
            \label{eq:Cs}
\end{customeq}
Other post-processing optimization formul\ae{} are possible if, for example, one wishes to trade space for time and attempt true simultaneous contractions.  We have not taken that route, in part for simplicity, and in part because the concurrent element can be delegated to software libraries such as the Cyclops Tensor Framework~\cite{CTF}.

The memory-conservation function $\Cspace{\cdot}$ is realized in  \cref{alg:mem-seq}, which constructs a concomitant concrete contraction sequence and returns it as a list of edges. If $\wpar{\cdot}$ and $\cspaceI{\cdot}$ are appropriately precomputed, the time complexity of \cref{alg:mem-seq} remains $\Th{\n}$, which leaves \ratcon at $\O{\n^4 \cdot \log\BsI{\G}}$. From the output sequence we may calculate $\Ctimename$ in linear time.

\subsection{Hardware}
All tests have been conducted on AWS EC2 \texttt{c5.large} instances running Ubuntu 16.04~\cite{aws}.  \Ratcon-related processes are implemented in Python\ 3.6 and byte-compiled using PyPy\,3.6\ v7.1.0, approximately doubling their performance as compared to basic CPython.  %
  For \Netcon, we use the authors' original C++ backend~\cite{Netcon}, compiled through GCC\ 5.4.0 with \emph{-O3} optimization.  There is also a MATLAB frontend which we have adapted to GNU Octave, but the C++ component accounts for more than 98\% of the execution time. Our implementation of the \ratcatcher incorporates some, but not all, of the optimizations described in the $A_1$ implementation by \textcite{BianRes}.

\newcommand{\colheadf}[1]{\texttt{#1}}
\newcommand{\cwcol}{$\cw{}$\xspace}
\newcommand{\cwtimecol}{\colheadf{$\cw{}$ time}\xspace}
\newcommand{\ectimecol}{\colheadf{Average EC time}\xspace}
\newcommand{\ratcontimecol}{\colheadf{Ratcon time}\xspace}
\newcommand{\ratconCtcol}{\colheadf{Ratcon{} $\Ctimename$}\xspace}
\newcommand{\netconCtcol}{\colheadf{Netcon{} $\Ctimename$}\xspace}
\newcommand{\netcontimecol}{\colheadf{Netcon{} time}\xspace}
\newcommand{\timeratio}{\ensuremath{\frac{\ratcontimecol}{\netcontimecol}}\xspace}
\newcommand{\errorfactor}{\ensuremath{\rho}\xspace}
\newcommand{\Ctratiocol}{\errorfactor\xspace}
\newcommand{\timeratiocol}{\ensuremath{\tau}\xspace}
\newcommand{\Ctratio}{
  \ensuremath{\frac{\ratconCtcol{}}{\netconCtcol{}}%
}\xspace}%
\begin{table}[ht]
\centering
\subcaptionbox{%
Mean%
    \label{tab:mean-log}
}{%
\resizebox{\textwidth}{!}{%
\begin{tabular}{|c|r|r|r|r|r|r|r|r|r|}
\hline
$L$         & \cwcol    & \cwtimecol & \ectimecol & \ratcontimecol & \ratconCtcol & \netconCtcol & \netcontimecol & \timeratiocol & \Ctratiocol \\ \hline
10 & 33.16 &	1.18 &	2.33 &	232.76 &	4.45E+15 &	1.06E+14* &	4251.19* &	0.09* &	3.36*           \\ \hline
9  & 30.30 &	0.65 &	1.20 &	122.40 &	8.62E+13 &	2.02E+13* &	1176.04*	 & 0.34*	 & 12.28*          \\ \hline
8  & 34.42	& 0.44 &	0.59 &	59.77 &	4.83E+15 &	1.61E+15 &	147.33 &	1.71 &	3.24           \\ \hline
7  & 33.59 &	0.25 &	0.26 &	26.00 &	3.36E+15 &	1.41E+15 &	60.58 &	2.75 &	6.49           \\ \hline
6  & 36.77	& 0.14	& 0.10 &	10.64 &	5.48E+15 &	3.73E+15 &	1.42  &	18.69 &	3.92           \\ \hline
5  & 34.66 &	0.06 &	0.03 &	3.69 &	1.09E+15 &	5.82E+14 &	0.12 &	47.87 &	3.26          \\ \hline
\end{tabular}
}%
}%
\par\bigskip
\subcaptionbox{%
    Median%
    \label{tab:median-log}
}{%
    \resizebox{\textwidth}{!}{%
\begin{tabular}{|c|r|r|r|r|r|r|r|r|r|} \hline
$L$         & \cwcol    & \cwtimecol & \ectimecol & \ratcontimecol & \ratconCtcol & \netconCtcol & \netcontimecol & \timeratiocol & \Ctratiocol \\ \hline
10 & 33.31 &	1.18 &	2.26 &	232.91 &	6.24E+14 &	3.30E+13* &	4987.61* &	0.05* &	2.74* \\ \hline
9 & 30.76 &	0.61 &	1.22 &	120.28 &	2.76E+13 &	7.52E+12* &	644.79* &	0.18* &	3.11* \\ \hline
8 & 34.53 &	0.43 &	0.60 &	59.80 &	1.33E+15 &	4.61E+14 &	69.38 &	0.85 &	2.44 \\ \hline
7 & 33.30 &	0.24 &	0.24 &	25.42 &	1.51E+14 &	5.34E+13 &	23.65 &	1.04 &	2.10 \\ \hline
6 & 36.80 &	0.12 &	0.10 &	10.44 &	2.39E+15 &	1.46E+15 &	0.72 &	15.42 &	1.74 \\ \hline
5 & 34.54 &	0.05 &	0.03 &	3.52 &	1.30E+14 &	8.70E+13 &	0.11 &	35.11 &	1.55 \\ \hline
\end{tabular}
}%
}%
\par\bigskip
\subcaptionbox{%
    Standard deviation%
    \label{tab:stddev-log}
}{%
    \resizebox{\textwidth}{!}{%
\centering %
\begin{tabular}{|c|r|r|r|r|r|r|r|r|r|}
\hline
$L$         & \cwcol    & \cwtimecol & \ectimecol & \ratcontimecol & \ratconCtcol & \netconCtcol & \netcontimecol & \timeratiocol & \Ctratiocol \\ \hline
10 & 2.70 &	0.12 &	0.22 &	11.16 &	8.43E+15 &	2.22E+14* &	2009.80* &	0.10* &	1.48*   \\ \hline
9  & 2.34 &	0.10 &	0.11 &	7.73 &	1.51E+14 &	2.91E+13* &	1425.22* &	0.33* &	44.85* \\ \hline
8  & 2.46 &	0.09 & 	0.07 &	3.46 &	7.71E+15 &	2.19E+15 &	230.54 &	1.93 &	2.61   \\ \hline
7  & 2.92 &	0.08 &	0.05 &	2.20 &	1.10E+16 &	4.53E+15 &	140.04 &	4.16 &	9.52  \\ \hline
6  & 1.60 &	0.05 &	0.01 &	1.26 &	6.55E+15 &	5.52E+15 &	1.78 &	16.62 &	6.64   \\ \hline
5  & 2.60 &	0.04 &	0.00 &	0.88 &	1.99E+15 &	9.21E+14 &	0.08 &	33.83 &	3.74   \\ \hline
\end{tabular}
}%
}%
\caption{Results.  All times presented are in seconds.  %
$\timeratiocol = \timeratio$, $\Ctratiocol = \Ctratio$.}
\label{tab:results}
\end{table}

\subsection{Experimental results} \label{sec:Results}
\Cref{tab:results} summarizes the statistics after running both \Netcon and \Ratcon on $30$ sample graphs per $L$ value, with edge-contraction sample size $N=100$, for $L\in [5..10]$.  
We use this range simply because for $L<5$ the single-term optimization problem is too easy, and for $L>10$ \Netcon is too slow.  

The `\cwtimecol' column records the time spent by the \Ratcatcher solving the initial pseudopolynomial function problem, %
i.e., finding $\carw{G}=\log_2\BsI{G}$.
`\ratcontimecol' is this value plus the cumulative EC (edge contraction) time;  `\ectimecol' is the mean iteration time thereof.  `\ratconCtcol' reflects the best \ctrees found for each graph.

For each graph \Netcon was used to find the optimum $\Ctimename$, unless its running time, given in the `\netcontimecol' column, exceeded a predefined limit of 7200 seconds, or two hours, in which case it got prematurely terminated.  This occurred for one $L=9$ sample and fully sixteen out of thirty $L=10$ graphs as \Netcon reached its operational limits.  Cells in the tables marked with an asterisk had these incomplete samples excluded from their calculation.

\timeratiocol is the ratio of the running times on a graph-by-graph basis.  \Ratcon underperforms for small $L$, where the time lost in repeated ECs dominates, then becomes exponentially faster at $L=9$ or $10$.  
The last column, %
\Ctratiocol, is the error factor in the total time estimation.  
For example, the mean $3.92$ for $L=6$ indicates that if a full tensor-network contraction were performed on one of these 36-vertex graphs, it could be expected to take four times as long using a \Ratcon-derived contraction order as the best possible.  The $L=9$ mean is skewed, by an order of magnitude, due to a single outlier boasting %
an error factor of $249$.  We never obtained an $\errorfactor{}$ less than $30$ for this particular graph, even after trying thousands of edge-contractions; we have to infer that this is one of the graphs for which the $\Bspacename$ approximation seldom or never results in a fast contraction order.%

\section{Conclusions and directions for future work} 
\label{sec:conclusion}
One way of applying these results is as follows.  A tensor contraction requires taking an equal number of sums as products, nearly; for complex numbers, addition and multiplication use 2 and 6 floating-point operations, respectively; therefore $8\cdot \Ct{\G}$ gives a good lower bound on the number of arithmetic instructions to run a full contraction sequence.  %
Referring to the statistics for the (completed) $L=10$ experiments, we can say that if a contraction sequence takes $0.09$ times as long to find with \Ratcon but $3.36$ times as long to execute, then the tipping-even point is found by
\renewcommand{\r}{\text{FLOPS}}
\begin{align*}
  \p{3.36 - 1}\cdot \frac{8\times 1.06\cdot 10^{14}}{\r} &\le  \p{1-0.09}\times %
  4.25\cdot 10^{3} \\
   \r &\ge 5.17\cdot 10^{11}\,,
\end{align*}
suggesting that \Ratcon would be the more efficient on a computer capable of 500\,GFLOPS or better.   %
In practice, we would recommend that the researcher always hedge her bets by running a \Netcon search in parallel with contraction of a \Ratcon-derived sequence.  If \Netcon comes up with a stronger sequence, she can restart the contraction and be none the worse.

Note that the times given are not intended as true benchmarks.  Their purpose is mainly to demonstrate how the relative error remains small and at the same order of magnitude as grid size increases.  Our \ratcatcher implementation is not especially heavily optimized and would be faster if it were rewritten in a statically-typed language such as C++, and included the rest of the enhancements proposed in~\cite{BianRes,HicksRat,HicksCycle}.  The heuristic from \cref{alg:mem-seq} could be replaced by the algorithm from \cite{Memory-optimal-evaluation}.

The greatest shortcoming of the \ratcatcher, obviously, is its planarity limitation.  Extending $\Bspacename$-optimization to general tensor networks would mean introducing another degree of approximation.  \textcite{Khuller&c}, for example, provide for an error factor of $\O{\log \n}$ on the carving-width; however, in moving to the multiplicative ring, this would translate to a space bottleneck of $\p{\Bspacename}^{\O{\log\n}}$.

The exponential bloat would apply equally if we preferred to switch to treewidth.  $\Btimename$ is intuitively closer than $\Bspacename$ to $\Ctimename$, and \textcite{ConSequence} report some (exact!) treewidth algorithms outperforming \Netcon in speed.  As previously mentioned, these apply only to \emph{unweighted} graphs; we would need to adapt them to weighted treewidth~\cite{WeightedTreewidth}, then use some form of the edge-contraction process to form a \ctree.  

Finally, inasmuch as the primary goal of this paper is to highlight the \ctree datatype itself, there is no reason not to apply ML or other metaheuristic techniques %
directly to the construction of these trees.  We have made a step in this direction with a simple genetic algorithm for evolving contraction sequences, included with the rest of the \Ratcon source code, which may be found at \url{https://github.com/TensorCon}.
\section*{Acknowledgments}
We would like to thank Illya Hicks for letting us use his \ratcatcher source code as a reference point for our own implementation; Robert Pfeifer and the American Physical Society for giving permission to redistribute \Netcon; and Eduardo Mucciolo for sharing the graphics of \cref{fig:half_sandwich,fig:sandwich}, as well as the term `bottleneck.'

\emph{This work was supported by National Science Foundation grant CCF-1525943.}
\printbibliography
\end{document}